\documentclass[reprint,superscriptaddress,amsmath,amssymb, aps,pra,prb,rmp,prstab,prstper,floatfix]{revtex4-2}
\usepackage{graphicx}
\usepackage{dcolumn}
\usepackage{bm}
\usepackage[mathlines]{lineno}
\usepackage{hyperref}
\hypersetup{colorlinks=true,linkcolor=red,citecolor=blue,urlcolor=magenta}
\usepackage{graphicx}
\usepackage{multibib}
\usepackage{braket}
\usepackage{dcolumn}
\usepackage{color}
\usepackage{times}
\usepackage{bm}
\usepackage{amssymb}
\usepackage{amsmath}
\usepackage{epsfig}
\usepackage{epstopdf}
\usepackage{dsfont}
\usepackage{subfigure}
\usepackage{verbatim}
\usepackage{verbatim}
\usepackage{braket}
\usepackage{float}
\usepackage{xcolor}
\definecolor{yel}{RGB}{255,0,0}
\begin{document}
	
	\title{Discrimination of Chiral Molecules through Holonomic Quantum Coherent Control}
	
	\author{Teng Liu}
	\thanks{These authors contributed equally}
	\affiliation{School of Physics \& Astronomy, Sun Yat-sen University, Zhuhai, Guangdong, 519082, China}
	
	\author{Fa Zhao}
	\thanks{These authors contributed equally}
	\affiliation{School of Physics \& Astronomy, Sun Yat-sen University, Zhuhai, Guangdong, 519082, China}
	
	\author{Pengfei Lu}
	\affiliation{School of Physics \& Astronomy, Sun Yat-sen University, Zhuhai, Guangdong, 519082, China}
	
	\author{Qifeng Lao}
	\affiliation{School of Physics \& Astronomy, Sun Yat-sen University, Zhuhai, Guangdong, 519082, China}
	
	\author{Min Ding}
	\affiliation{School of Physics \& Astronomy, Sun Yat-sen University, Zhuhai, Guangdong, 519082, China}
	
	\author{Ji Bian}
	\affiliation{School of Physics \& Astronomy, Sun Yat-sen University, Zhuhai, Guangdong, 519082, China}
	
	\author{Feng Zhu}
	\affiliation{School of Physics \& Astronomy, Sun Yat-sen University, Zhuhai, Guangdong, 519082, China}
	\affiliation{Shenzhen Research Institute of Sun Yat-Sen University, Nanshan Shenzhen 518087, China}
	
	\author{Le Luo}\email[]{luole5@mail.sysu.edu.cn}
	\affiliation{School of Physics \& Astronomy, Sun Yat-sen University, Zhuhai, Guangdong, 519082, China}
	\affiliation{Shenzhen Research Institute of Sun Yat-Sen University, Nanshan Shenzhen 518087, China}
	\affiliation{State Key Laboratory of Optoelectronic Materials and Technologies, Sun Yat-Sen University, Guangzhou 510275, China}
	\affiliation{International Quantum Academy, Shenzhen, 518048, China}

	\begin{abstract}
		A novel optical method for distinguishing chiral molecules is proposed and validated within a quantum simulator employing a trapped-ion qudit. This approach correlates the sign disparity of the dipole moment of chiral molecules with distinct cyclic evolution trajectories, yielding the unity population contrast induced by the different non-Abelian holonomies corresponding to the chirality. Harnessing the principles of holonomic quantum computation (HQC), our method achieves highly efficient, non-adiabatic, and robust detection and separation of chiral molecules.  Demonstrated in a trapped ion quantum simulator, this scheme achieves nearly 100\% contrast between the two enantiomers in the population of a specific state, showcasing its resilience to the noise inherent in the driving field.
	\end{abstract}
	
	\maketitle
	
\begin{figure}
		\centering
		\includegraphics[width=9cm]{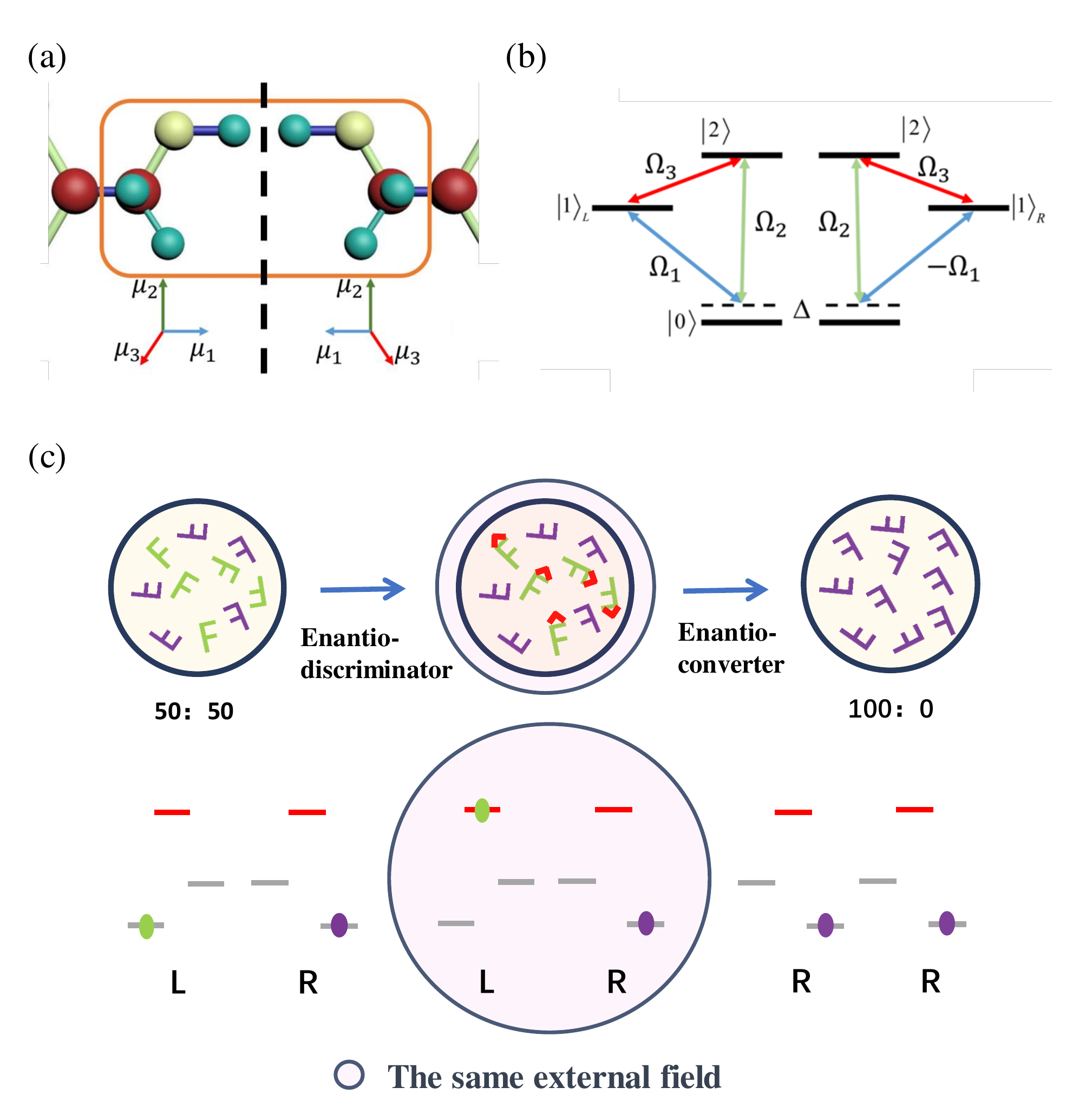}
		\caption{ (a) Schematic diagram of chiral cyclohexylmethanol molecule's partial structure. The difference of the two enantiomers can be reduced to the discrepancy of electric dipole moments $\mu_i$ in one of the three directions, here is $\mu_1$. (b) The electric dipole interaction model of chiral molecules with three different energy levels. $L$- and $R$-handedness share same coupling with ${\Omega _2}$, ${\Omega _3}$ and opposite sign of coupling ${\Omega _1}$. (c) Schematic of two-step asymmetric synthesis based on quantum coherent control.}
		\label{Fig.1}
	\end{figure}

	\textit{Introduction.}—Since Pasteur's discovery of chirality in his graceful tartaric acid experiment \cite{pasteur1848relations,gal2017pasteur}, omnipresent chiral molecules have been realized and have profoundly influenced many fields, including chemistry, biochemistry, pharmacology, and materials science. A chiral molecule, as known as an enantiomer, refers to the one that can overlap with its counterpart with the transformation of mirror symmetry (cyclohexylmethanol molecules shown in Fig. \ref{Fig.1} (a)). Two enantiomers usually share numerous physical properties like density and viscosity\cite{shubert2016chiral,quack2008high}, but the significant differences in chirality could emerge. An enantiomer drug (like R-thalidomide ) may be a fairly efficient medicament, while its counterpart (like S-thalidomide) may cut no ice or even result in detrimental reactions for living organisms \cite{eriksson2001clinical,teo2004clinical}. Therefore, it is imperative to differentiate enantiomers quickly and accurately.
	
	The early-stage methods\cite{ahuja2011chiral} for chiral molecule detection including crystallization, derivatization, kinetic resolution are typically complicated, expensive and laborious. 
	Alternatively, optical methods, such as optical rotary dispersion \cite{berova2011comprehensive,fischer2005nonlinear}, circular dichroism \cite{nafie1976vibrational,berova2000circular,nafie2011vibrational,lux2012circular,lehmann2013imaging,goetz2017theoretical,stephens1985theory}, and Raman optical activity \cite{barron2009molecular}, offer advantages in terms of simplicity and convenience, and are widely applied. The differential optical signal for the chiral molecules mainly originates from weak magnetic dipole or electric quadrupole interactions. Hence a variety of strategies for enhancing the optical signals have been developed, such as enhanced strong anti-Stokes Raman-scattering field\cite{begzjav2019enhanced}, circularly polarized X-ray light\cite{cireasa2015probing}, plasmonic metamaterials\cite{zhao2017chirality}, and various microwave-driven coherent population transfer techniques \cite{domingos2018sensing,koumarianou2022assignment,eibenberger2017enantiomer,perez2017coherent,cai2022enantiodetection}.

One notable approach to enhancing the signals is based on quantum coherent control (QCC) techniques, where chiral molecules are differentiated by precisely controlling the phases of external optical fields, allowing the quantum states of the molecules to evolve into completely different states. The most typical scheme is enantio-selective cyclic population transfer (CPT), proposed by Shapiro et al. in Ref. \cite{kral2001cyclic}. In this scheme, three optical fields are applied to couple three levels, respectively. The state evolution paths are separated due to the contrary signs of transition dipoles in the mirror-symmetric configurations, as shown in Figs. \ref{Fig.1}(a) and \ref{Fig.1}(b). Subsequently, based on CPT, a two-step asymmetric synthesis scheme is proposed \cite{kral2003two,gerbasi2004theory}, demonstrating the significant potential of QCC approaches not only for chiral discrimination but also for chiral purification and asymmetric synthesis. According to this design, one type of chiral molecule is excited through the CPT process, and subsequently, these ``marked" molecules undergo a coherent process to achieve conversion. This process is summarized in Fig. \ref{Fig.1}(c). Despite the requirement of molecular-scale quantum coherence time to be as long as possible in their adiabatic processes, these two schemes shed light on consecutive QCC methods \cite{gonzalez2001separation,kral2003two,thanopulos2003theory,li2008dynamic,vitanov2019highly,torosov2020efficient,torosov2020efficient,wu2020two,ding2022chiral,guo2022cyclic,ye2020fast,wu2019robust}.
The central issue of QCC approaches lies in how to rapidly and robustly induce molecules of different chirality to distinct energy levels under the same external field. Most existing QCC methods mentioned above are time-consuming and strongly depend on precise experimental control, lacking optimization for robustness against experimental noise, thus limiting the feasibility of their experimental implementation.

	In this work, we present a fast, robust and fewer pulse modulated high selectivity scheme using the method of geometric coherent control techniques, referred as Geometric QCC (GQCC), and validate it  with a qudit of trapped ${}^{171}Y{b^ + }$ ion. Firstly, we correlate the different signs of the dipole moment of chiral molecules with different geometric cyclic trajectories, constructing chiral-dependent quantum holonomies. Molecules with different chirality can thus be induced to highly distinguishable orthogonal final states under the same external fields. The geometric cyclic evolution and the process of geometric phase accumulation significantly enhance the robustness of our scheme to local control errors\cite{leibfried2003experimental,lupo2009robustness}. Second, we map the energy level structure of chiral molecules onto a qudit in a trapped ${}^{171}Y{b^ + }$ ion and demonstrate the robustness and selectivity of our scheme via a experiment of quantum simulation. Our work, for the first time, introduces the geometric coherent control techniques into chiral molecule discrimination. It not only explores the feasibility of QCC approaches in the realms of chiral discrimination and asymmetric synthesis but also provides a valuable paradigm for exploring quantum simulation and quantum control techniques in ion traps applied to specific chemical processes.	
	
	\textit{Holonomic coherent control for chiral molecules.}—In the general CPT based scheme, the closed-loop population transfer exhibits phase sensitivity, as their phase has a $\pi$ difference for different enantiomers due to the opposite transition dipoles. In this process, three overlapped and  modulated resonate pulses are required. We propose that the opposite transition dipoles can also be mapped onto different pure geometrical quantum holonomy targeting  $\ket{2}$ and $\ket{1_L}$($\ket{1_R}$) subspaces. Population can be naturally induced onto the opposite levels by different paths. Our scheme relaxes the constraints of the three-pulse joint modulation, obtaining more robust and higher contrast chiral discrimination

	Our path design is grounded in quantum holonomy theory\cite{aharonov1987phase}. For an initial state $|\psi(0)\rangle$ following the Schrödinger equation, we can represent the final states as $|\psi(t)\rangle=\mathcal{T} e^{-i \int_0^t H(\tau_0) d \tau_0}|\psi(0)\rangle$, where $\mathcal{T}$ is the time ordering operator and $\hbar=1$ for simplicity. At each moment, we can define a complete set of basis \{$\ket{\zeta_m(t)}$\}, therefore the states can be represented as 
	$
		|\psi(t)\rangle=\sum_m \alpha_m(t)\left|\zeta_m(t)\right\rangle.
	$
	The time-dependent bases $\ket{\zeta_m(t)}$ characterizes the local geometric properties of the evolution trajectory at each moment. When $t=T$, the state vector has completed a cyclic evolution along closed trajectories with $\zeta_m(T)=\zeta_m(0)$, and the final state is
	$
	|\psi(T)\rangle=\sum_m \alpha_m(T)\left|\zeta_m(0)\right\rangle.
    $
    Compare to the initial state
     $|\psi(0)\rangle=\sum_m \alpha_m(0)\left|\zeta_m(0)\right\rangle$,
    $\alpha_m(T)$ contains the geometric information of the cyclic evolution path. The dynamics of $\alpha_m(t)$ follow:
	\begin{equation}
		\frac{d}{d t} \alpha_m(t)=i \sum_n\left[A_{m n}(t)-H_{m n}(t)\right] \alpha_n(t)
		\label{alpha}
	\end{equation}
	where 
	$H_{m n}(t)=\left\langle\zeta_m(t)|H(t)| \zeta_n(t)\right\rangle$ is the element of dynamical part matrix $\boldsymbol{H}$, $A_{m n}(t)=i\left\langle\zeta_m(t)|(d / d t)| \zeta_n(t)\right\rangle$ is element of geometric part $\boldsymbol{A}$ which depends on the change of bases owing to the local curvature. Through appropriate parameter configurations, the dynamical part can be set to zero, thereby achieving purely geometric quantum evolution. 
	
In the following, we construct two holonomies for different enantiomers, which transfer the sign of transition dipoles of different chiral molecules onto two different bases. For the three levels in Fig. \ref{Fig.1}(b), we drive the time-dependent couplings ${\Omega _1}(t)  = \Omega (t)\sin \frac{\theta }{2}{e^{i\Phi (t)}}$ between $\ket{0}$ and $\ket{1}$ and ${\Omega _2}(t) = \Omega (t)\cos \frac{\theta }{2}{e^{i\left( {\Phi (t) + \phi } \right)}}$, then incorporate the different transition dipole moment signs into the Hamiltonian,
		\begin{equation}
				{H_{L/R}}(t){\rm{ }} =\left(
				\begin{array}{ccc}
					0 & 0 & \pm\Omega_1(t)/2\\
					0 & 0 & \Omega_2(t)/2\\
				\pm\Omega_1(t)^*/2 & \Omega_2(t)^* /2& \Delta(t) \\
				\end{array}
				\right)
			\label{Eq.4}
		\end{equation}
	where $\Delta (t)$, $\Omega(t)$ and $\Phi(t)$ are modulated detuning, Rabi frequency and phase respectively (In supplementary material S1 for more details). The "+" represents $L$-handedness and "-" represents $R$-handedness here. According to Eq. (\ref{Eq.4}), we can define the basis in dark and bright state space with $\ket{D_L} = -\cos (\theta / 2)|1\rangle+\sin (\theta / 2) e^{i\phi}|2\rangle$, $\ket{D_R} = \cos (\theta / 2)|1\rangle+\sin (\theta / 2) e^{i\phi}|2\rangle$, and $\ket{ B_L}=\sin(\theta/2)\ket{1}+\cos(\theta/2)e^{i\phi}\ket{2}$, $\ket{ B_R}=-\sin(\theta/2)\ket{1}+\cos(\theta/2)e^{i\phi}\ket{2}$.
	Thus Eq. (\ref{Eq.4}) can be reduce to 
	\begin{equation}
		H_{L/R}(t)=\Delta(t)|0\rangle\langle 0|+\left(\frac{\Omega(t)}{2} e^{i \Phi(t)}\left|B_{L / R}\right\rangle\langle 0|+\text { H.c. }\right)\end{equation}
	
\begin{figure}
		\centering
		\includegraphics[width=9cm]{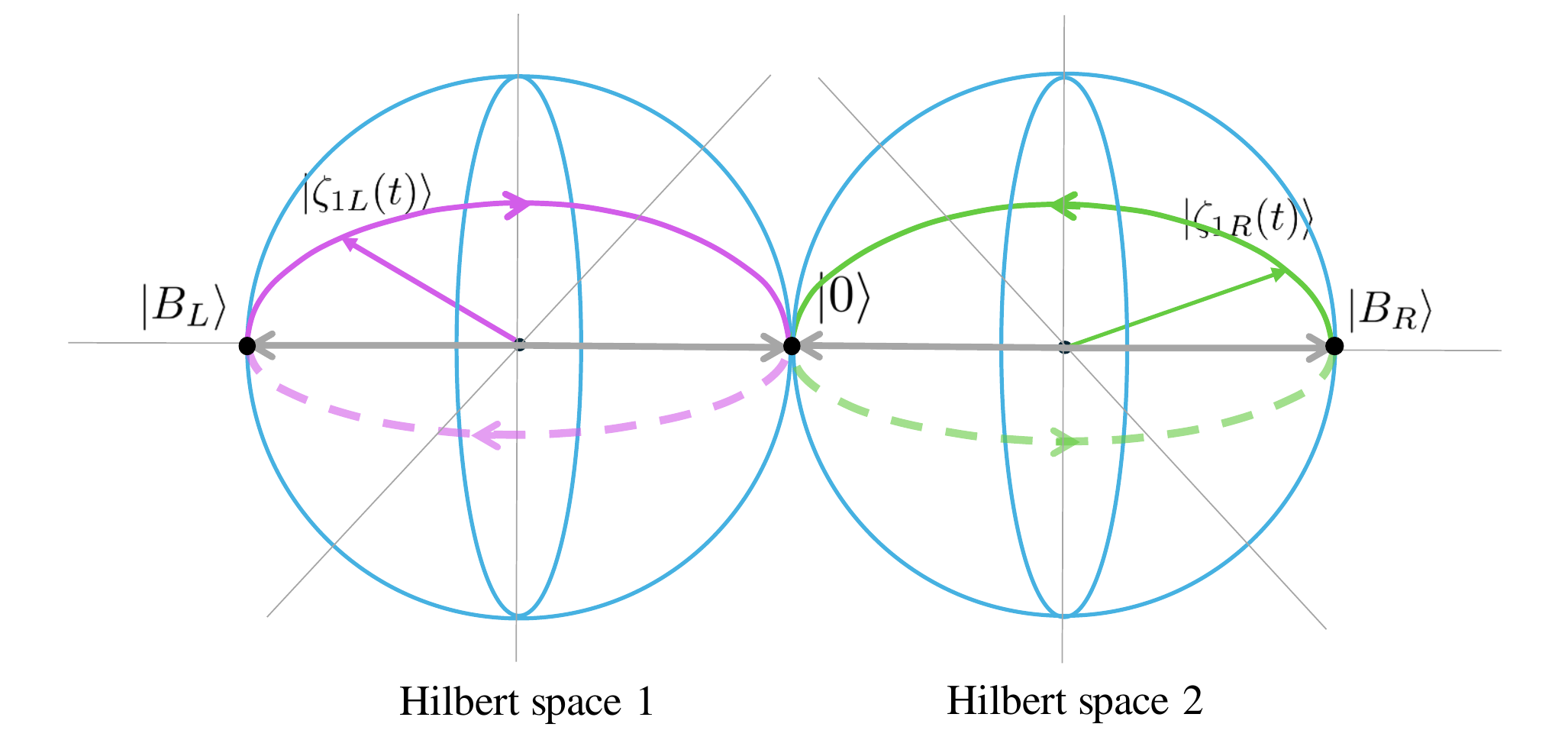}
		\caption{Schematic of the cyclic trajectory of basis $\ket{ \zeta_{1L}(t)}$ and $\ket{ \zeta_{1R}(t)}$ in Eq. (\ref{psiL}). Purple line represents $L$-handedness, and green represents $R$-handedness. Hilbert space 1 and 2 are spanned by $\ket{B_L}$, $\ket{0}$ and $\ket{B_R}$, $\ket{0}$ respectively. }
		\label{holonomy}
	\end{figure}

	Within this framework, the bright state $\ket{ B_{L/R}}$ couples with $\ket{0}$ while the dark state $\ket{ D_{L/R}}$ is decoupled. We can define the invariant bases $\ket{\zeta_{0L}}=\ket{D_L}$, $\ket{\zeta_{0R}}=\ket{D_R}$, and the time-dependent bases along the trajectories of cyclic evolution for different handedness,
	\begin{equation}
		\ket{ \zeta_{1L/R}(t)}=\sin \frac{k(t)}{2}|B_{L/R}\rangle+\cos \frac{k(t)}{2} e^{-i \beta(t)}|0\rangle
		\label{psiL}
	\end{equation}
	where $k(t)$ and $\beta(t)$ indicate the characteristics of the cyclic path (In supplementary material S1 for details). The bases in Eq. (\ref{psiL}) satisfy the cyclic evolution condition $\ket{ \zeta_{1L/R}(0)}=\ket{ \zeta_{1L/R}(T)}$ with $k(T)=k(0)=\pi$, where $T$ is the cyclic evolution time, shown in Fig. \ref{holonomy}. Substitute Eq. (\ref{psiL}) into Eq. (\ref{alpha}), if condition for diagonalization of matrix $\boldsymbol{A}-\boldsymbol{H}$ is satisfied, we can obtain $\alpha_m(t)=e^{i \int_0^t [A_{mm}(\tau)-H_{mm}(\tau)] d \tau}$. At the end of cyclic evolution, if the dynamic phase $\int_0^t H_{mn}=0$, the time dependent bases $\ket{ \zeta_{1L/R}(0)}$ will acquire a pure geometric phase $\gamma=\int_0^t i\left\langle\zeta_{1L/R}(\tau)|(d / d \tau)| \zeta_{1L/R}(\tau)\right\rangle d\tau$, and the path-dependent evolution operator can thus be written as:
	\begin{equation}
		U(T, 0)_{L/R}=\left|\zeta_{0L/R}\right\rangle\bra{\zeta_{0L/R}}+e^{i \gamma}\ket{ \zeta_{1L/R}(0)}\left\langle\zeta_{1L/R}(0)\right|,
		\label{UL}
	\end{equation}

	Herein, we construct distinct quantum holonomies for different chirality of chiral molecules.	
	By assigning suitable values to the parameters, for instance, we set parameters $\theta  =  \frac{3\pi }{4}$, and $\gamma  =  \pi$, we can express Eq. (\ref{UL}) in the following form
	\begin{equation}
		U(T, 0)_{L/R}=\left(\begin{array}{ccc}
			\frac{1}{\sqrt{2}} & \pm \frac{1}{\sqrt{2}} e^{-i \phi} & 0 \\
			\pm \frac{1}{\sqrt{2}} e^{i \phi} & -\frac{1}{\sqrt{2}} & 0 \\
			0 & 0 & \epsilon
		\end{array}\right),
		\label{Eq.1}
	\end{equation}
	 where $\epsilon$ is the unity complex number, which ensures the unitarity of the $U(\tau, 0)_{L/R}$, "$\pm$" present different signs of $\Omega_1(t)$. In this process, we prepare the initial state of to chiral-independent form
	$\left|\psi_{initial}\right\rangle=\frac{1}{\sqrt{2}}\left(|1\rangle+e^{i \varphi_0}|2\rangle\right)$, 
	where the relative phase $\varphi_0$ can be arbitrary as the distinguish signals can be found by scanning the phase $\phi$ to $\phi=\varphi_0$ in Eq. (\ref{Eq.1}). For $L$-handedness molecules, the final state will be transferred to $\left|\psi_{final}\right\rangle_L=\ket{1}$, and for $R$-handedness $\left|\psi_{final}\right\rangle_R=\ket{2}$. We can choose to detect either chiral state $\ket{1}$ or non-chiral state $\ket{2}$, both of which shows 100\% population contrast theoretically.
	
    The requirement of accumulating pure geometric phases in Eq. (\ref{UL}) is necessary in our sheme, while the decoupling of dynamical phases maximally reduces the impact of control errors such as Rabi errors, enhancing the robustness of our scheme significantly\cite{thomas2011robustness}. However, solving this problem usually requires strict parameter restrictions and complex pulse optimization. Recently, relevant solutions have been used to tackle problems of constructing holonomic quantum computation (HQC) in multi-level subspaces \cite{sjoqvist2012non,xu2012nonadiabatic,liu2019plug}. To meet the constraints of coherent time and improve detection efficiency, we focus on the principle of non-adiabatic scheme NHQC in Ref.\cite{sjoqvist2012non} and a more flexible and fewer-parameter-constraints scheme NHQC+ in Ref.\cite{liu2019plug}. In NHQC based scheme, two resonance pulses ($\Delta(t)=0$) are applied between $\ket{0}$ and $\ket{1}$, $\ket{0}$ and $\ket{2}$. The condition $\left\langle {{\zeta _m}(t)} \right|H(t)\left| {{\zeta _n}(t)} \right\rangle  = 0$ is satisfied at each moment\cite{sjoqvist2012non,xu2012nonadiabatic}. In NHQC+ based scheme
	, non-resonant field is applied ($\Delta(t)\neq0$), and the auxiliary basis are choose to satisfy the initial condition $\langle {\zeta _m}(0)|{H_{eff}}(t)\left| {{\zeta _n}(0)} \right\rangle  = 0$ and pure geometric condition $\int_0^\tau  {\langle {\mu _m}(t)|H(t)\left| {{\mu _n}(t)} \right\rangle } dt = 0$ \cite{liu2019plug}.
	The details of our parameters setting based on NHQC and NHQC+ can be found in supplementary material S1.
	
\begin{figure*}
    	\centering
    	\includegraphics[width=18cm]{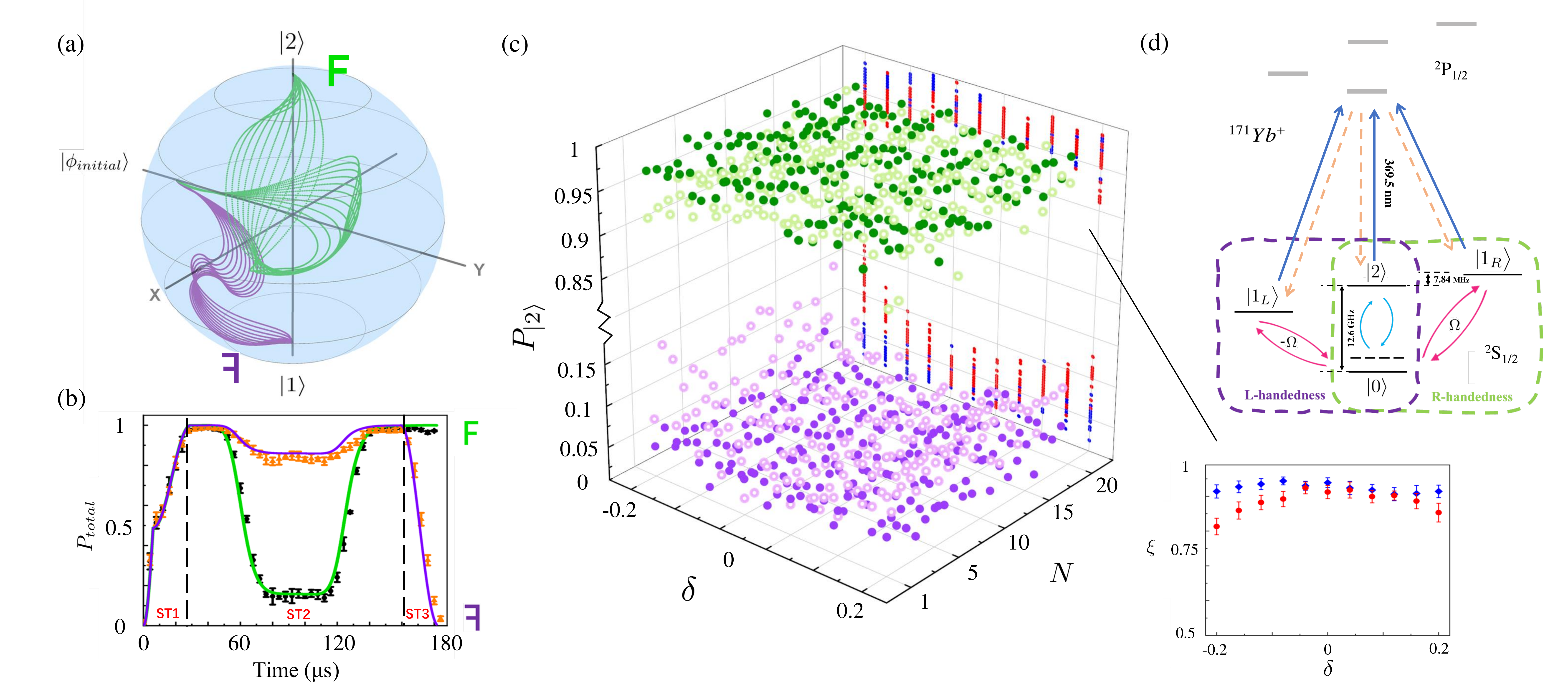}
    	\caption{Experimental results for separating chiral molecules. (a) The evolution trajectories of different chiral molecules on the Bloch sphere. the purple and green line are represent $R$-handedness and $L$-handedness respectively. They are prepared in the same initial state $\ket{\phi_{initial}}=(\ket{2}+\ket{1_R}(\ket{1_L}))/\sqrt{2}$, and then induced into completely different final states $\ket{2}$ or $\ket{1_R}(\ket{1_L})$ with robust geometric quantum evolution. Each colored trajectory contains 10 offset errors $\delta$. (b) The total population $P_{total}$ evolution. The experiment data plotted with orange (black) triangle (circle) and the theoretical line in purple and green are represent $R$-handedness and $L$-handedness respectively. (c) Experimental results of NHQC and NHQC+ schemes for distinguishing chiral molecules. The light purple and light green hollow circles represent the experimental results for NHQC, while the purple and green solid circles represent  NHQC+. The subplots illustrate the contrast $\xi$ of both schemes under different offset errors $\delta$. (d) The configuration of energy levels for chiral molecules. }
    	\label{experiment}
    \end{figure*}

	\textit{Quantum simulation through a trapped-ion qudit.}—
To validate the above theoretical scheme, we implement a quantum simulation through a trapped ion qudit. For a trapped $^{171}Yb^+$ ion, the hyperfine levels $\lvert 0 \big\rangle  = \lvert {{}^2{S_{1/2}},F = 0} \big\rangle $, $\lvert 2 \big\rangle  = \lvert {{}^2{S_{1/2}},F = 1,{m_F} = 0} \big\rangle $ and Zeeman levels $\lvert 1_R \big\rangle  = \lvert {{}^2{S_{1/2}},F = 1,{m_F} = 1} \big\rangle $, $\lvert 1_L \big\rangle  = \lvert {{}^2{S_{1/2}},F = 1,{m_F} = -1} \big\rangle $ constitute a highly controllable four-level qudit system as shown in Fig. \ref{experiment}(d). By applying a static magnetic field of 5.6 G in the Z-direction, the energy levels split by 7.84 MHz. The qubit can be encoded on $\lvert 1_L \big\rangle$($\lvert 1_R \big\rangle$) and $\lvert 2 \big\rangle$ subspace, which was later used for distinguishing chiral molecules and detection. Two amplitude, frequency and phase modulated microwaves are applied to drive the couple between  $\lvert 0 \big\rangle$, $\lvert 1_R \big\rangle$ and $\lvert 0 \big\rangle$, $\lvert 2 \big\rangle$. We test the $U_1(\pi/2, 0, \pi)$ and $U_2(3\pi/4, \pi, \pi)$ gates with fidelity up to $99.7\% \pm 0.17\%$ and $99.56\% \pm 0.21\%$ respectively. The quantum process tomography (QPT) is also implemented. The process fidelity of $U_1$ for NHQC+ scheme is $91.00\%$. The detail of experiment setup can be found in supplementary material S2.
	
	We have successfully validated our GQCC scheme for chiral molecule discrimination in our trapped ion quantum simulator. We choose $\lvert 0 \big\rangle$, $\lvert 1_L \big\rangle$, $\lvert 1_R \big\rangle$ and $\lvert 2 \big\rangle$ to establish the model of chiral molecules, as shown in Fig. \ref{experiment}(d). The states $\lvert 0 \big\rangle $, $\lvert 1_R \big\rangle $ and $\lvert 2 \big\rangle $ form the $R$-handedness molecule, and the state $\lvert 0 \big\rangle $, $\lvert 1_L \big\rangle $ and $\lvert 2 \big\rangle $ form the $L$-handedness molecule. The opposite signs of Rabi frequency is incorporated into our driving field waveform, which leads to similar chiral structures in Fig. \ref{Fig.1}(b). The state is initialized to superposition state $\lvert \psi  \big\rangle  =  \frac{1}{{\sqrt 2 }}\lvert 1_L \big\rangle \left( {\lvert 1_R \big\rangle } \right) + \frac{i}{{\sqrt 2 }}\lvert 2 \big\rangle $, and we detect the population on the state $\lvert 2 \big\rangle $ at the end of external field interaction. In Fig. \ref{experiment}(b), we show the $P_{total}$ evolution of both $L$- and $R$-handedness model, where $P_{total}=P_{\ket{1_L}}(P_{\ket{1_R}})+P_{\ket{2}}$ represents the total population of $\ket{1_R}$($\ket{1_L}$) and $\ket{2}$. At the detection stage ST3 in Fig. \ref{experiment}(b), we applying a $\pi$ pulse between $\ket{1_R}$($\ket{1_L}$) and $\ket{0}$, showing completely distinct population signals between $\ket{1_R}$($\ket{1_L}$) and $\ket{0}$ (Refer to supplementary material S2).

	We further explored the stability of our GQCC scheme in chiral molecule discrimination. We define the absolute difference between the population of left-handed and right-handed molecules on the $\lvert 0 \big\rangle $ as the discrimination contrast parameters $\xi=\lvert P_{\ket{2}}-P_{\ket{1_L}}(P_{\ket{1_R}})\lvert$, $0 \le \xi  \le 1$. $\xi=0$ means that chirality cannot be distinguished, while $\xi=1$ means complete discrimination. We added a set of offset errors $\delta = \lvert\Omega_r(t)-\Omega_i(t)\lvert/\Omega_i(t)$ to the driving field, where  ${\Omega_r}(t)$ stands for the pulse amplitudes in reality and ${\Omega_i}(t)$ stands for the amplitude in ideality. We choose 10 values of $\delta$ from $-20\%$ to $20\%$ with a step of $4\%$. For each $\delta$, we repeat the experiment $N =20$ times. The population of different handedness chiral molecule models are almost completely separated and both of our NHQC+ and NHQC based GQCC sechemes maintained a very high contrast under experiment errors, as shown in Fig. \ref{experiment}(c) and its subplot. In order to make this excellent robustness of our scheme more intuitive, we plot the evolution trajectory of quantum states in the Hilbert space under these errors in Fig. \ref{experiment}(a). Almost all trajectories ultimately evolve into two mutually orthogonal states that can be perfectly distinguished with high contrast.
	
  \textit{Robustness.}—Compared to general QCC chiral discrimination scheme, our approach exhibits high robustness and flexibility. Recently, STA-based scheme in Ref.\cite{vitanov2019highly} accomplished rapid chirality-dependent population transfer of certain energy levels. This scheme solves the issue of excessively long adiabatic evolution time in the original approach in Ref.\cite{kral2001cyclic}, since STA technique was initially developed to introduce an additional laser pulse in the stimulated Raman adiabatic passage process (STIRAP) to ensure that the system undergoes a equivalent perfect adiabatic evolution\cite{guery2019shortcuts}. However, in the design process of STA pulses, the primary emphasis has been on accelerating the transfer of quantum system populations, with less consideration given to its robustness. In our scheme, the pure geometric phase accumulation in Eq. (\ref{UL}) indicates that the final state mainly depends on the whole trajectory in parameter space and is insensitive to local variations, which significantly enhances robustness. Meanwhile, within HQC, the non-adiabatic quantum holonomy methods also enable our approach to overcome the time-consuming issue in original scheme.
  
\begin{figure}
	\centering
	\includegraphics[width=9cm]{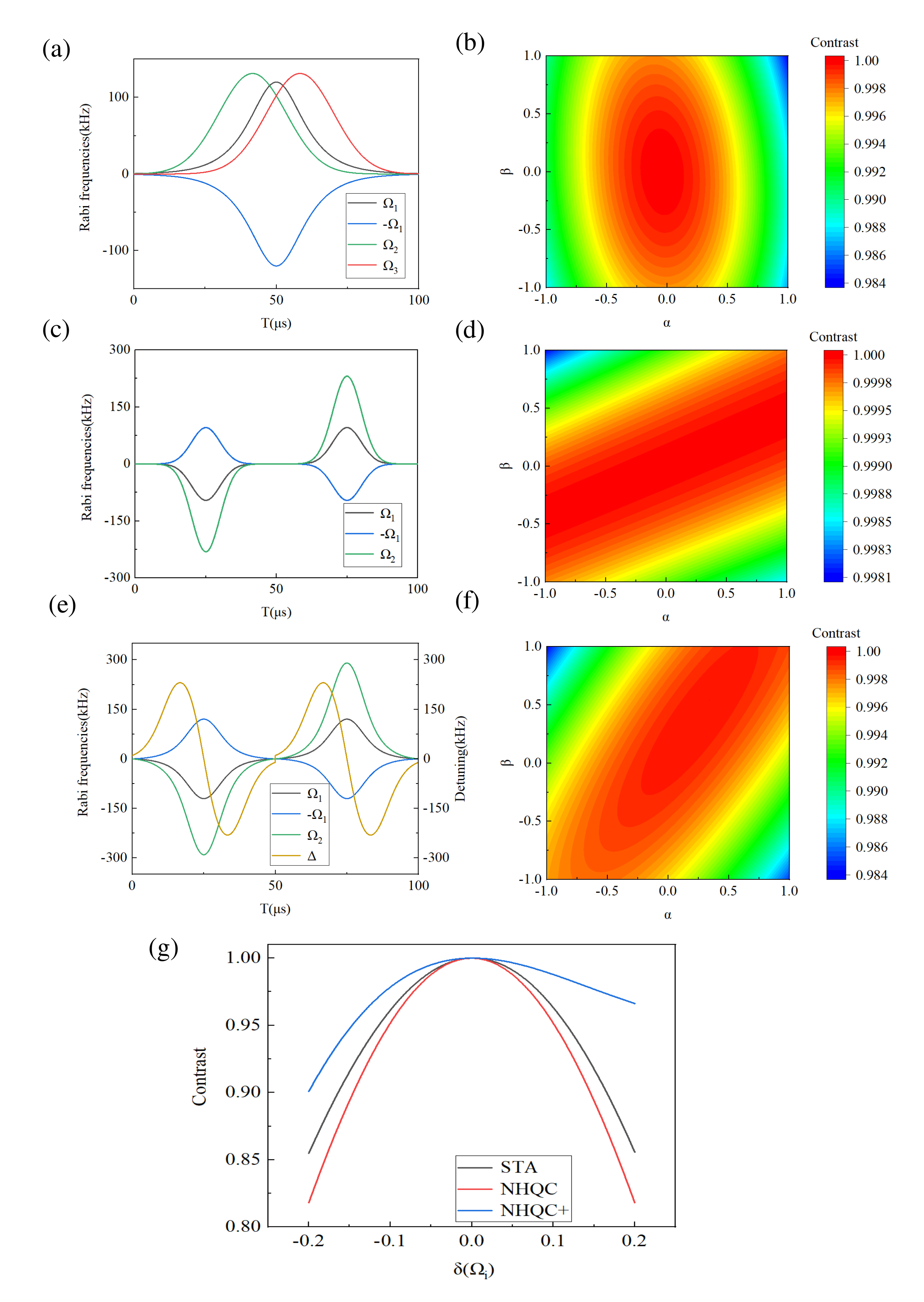}
	\caption{The influence of Rabi frequency errors on the final state contrast is examined for STA, NHQC, and NHQC+ schemes. Graphs (a), (c), and (e) respectively illustrate the experimental pulse profiles for the STA scheme, NHQC scheme, and NHQC+ scheme. In these graphs, the STA scheme and NHQC scheme operate under resonance conditions, while the NHQC+ scheme operates under detuned conditions. The final contrast of the different schemes under systematic errors is displayed in graph (g), while the final contrast under random errors is shown in graphs (b), (d), and (f).}
	\label{system error}
\end{figure}

	We evaluate the robustness of our approach against Rabi frequency random error and offset error, and compared it with the STA based scheme. The specific determination of parameters can be found in supplementary material S1. For the case of random fluctuations in the Rabi frequency, we analyze the impact of two noise models: random noise and Gaussian white noise. The affected Rabi frequency can be written as	${\Omega _r}(t) = {\Omega _i}(t)(1 + \alpha \cdot {\rm{awag}}(0.2) + \beta \cdot{\rm{rand}}(0.2))
		\label{Eq.10}$,
	where ${\rm{awag}}(0.2)$ means a white Gaussian noise with standard deviation 0.2, ${\rm{rand}}(0.2)$ means random function with lower bound -0.2 and upper bound 0.2, $\alpha$ and $\beta$ are proportional parameters. The performance of our schemes against system and random errors are shown in Figs. \ref{system error}(b)(d)(f)(g), and the corresponding pulse diagrams with same cost of time is illustrated in Fig. \ref{system error}(a)(b)(c). For the offset error, an relative deviation parameter $\delta = \lvert\Omega_r(t)-\Omega_i(t)\lvert/\Omega_i(t)$ is considered. We setting this parameter's range from -20\% to 20\%, and the corresponding discrimination contrasts are presented in Fig. \ref{system error}(g). The numerical simulation results of pulse shape and error in Fig. \ref{system error} indicate that our proposed scheme based on geometric coherent control exhibits stronger robustness to both types of errors and requires fewer modulated pulses.

\textit{Conclusion.}—
We have demonstrated a scheme for the discrimination of chiral molecules based on geometric coherent control, and we further experimentally simulate the discrimination and detection process of our scheme in a trapped ion qudit. By mapping the evolution path of different chiral molecules onto different non-adiabatic quantum holonomies, we not only overcome the requirement for adiabatic evolution in the general QCC method but also enhance the robustness of this process. Our approach requires fewer modulated fields and offers more flexible choices for detectable levels, demonstrating significant potential for applications in reality. Beyond chiral discrimination and asymmetric synthesis, geometric coherent control demonstrated in our paper could find emergent applications in the field of precision control of molecular reaction pathways and chemical quantum dynamics\cite{brumer1986control, koch2019quantum, brif2010control, judson1992teaching}. Moreover, we showcase the capability to utilize quantum computing machines, such as the trapped ion qudit in this work, for simulating complex chemical problems.

Thanks to Dr. Xueke Song, Dr. Dong Wang, Dr. Liu Ye, Dr. Baojie Liu for helpful discussion and Dr. Yang Liu,  Dr. Su Zhan, Xinxin Rao and Xiao Song for experimental support. 
\bibliographystyle{unsrt}
\bibliography{REF_DCM}
\end{document}